\documentclass[5p,10pt,twocolumn]{elsarticle}

\usepackage{amssymb}
\usepackage{epsfig}
\usepackage{url}
\usepackage{amsmath, amsthm, amssymb}
\usepackage{graphicx}
\usepackage{picins}

\newtheorem{Lemma}{Lemma}

\newtheorem{remark}{Remark}
\newtheorem{Theorem}{Theorem}

\usepackage{algorithm}
\usepackage{algorithmic}
\usepackage{multirow}
\usepackage{amstext}

\usepackage{url}
\usepackage{color}

\newcommand{\warn}[1]{}
\newcommand{\nop}[1]{}

\journal{Journal of Network and Computer Applications}

\begin{document}

\begin{frontmatter}



\title{Taking A Free Ride for Routing Topology Inference in Peer-to-Peer Networks}


\author[label1,label2]{Peng Qin}
\author[label1]{Bin Dai\corref{cor1}}
\author[label2]{Kui Wu}
\author[label1]{Benxiong Huang}
\author[label1]{Guan Xu}

\address[label1]{Department of Electronics and Information Engineering\\
Huazhong University of Science and Technology, Wuhan, China}
\address[label2]{Department of Computer Science\\
University of Victoria, Victoria, Canada}
\cortext[cor1]{Corresponding author. Tel./fax: +86 2787541604. \\
E-mail address: nease.dai@gmail.com (B. Dai).}

\begin{abstract}
A Peer-to-Peer (P2P) network can boost its performance if peers are provided with underlying network-layer routing topology. The task of inferring the network-layer routing topology and link performance from an end host to a set of other hosts is termed as network tomography, and it normally requires host computers to send probing messages. We design a passive network tomography method that does not requires any probing messages and takes a free ride over data flows in P2P networks. It infers routing topology based on end-to-end delay correlation estimation (DCE) without requiring any synchronization or cooperation from the intermediate routers. We implement and test our method in the real world Internet environment and achieved the accuracy of $92\%$ in topology recovery. We also perform extensive simulation in OMNet++ to evaluate its performance over large scale networks, showing that its topology recovery accuracy is about $95\%$ for large networks.

\end{abstract}


\end{frontmatter}


\section{Introduction}
\label{sec::introduction}
\vspace{-0.1cm}

Peer-to-Peer (P2P) networks have been broadly adopted for large-scale content distribution and online multimedia streaming~\cite{Stephanos::SurP2PConDisTec::2004,Dongyu::ModPerAnaBTP2PNet::2004,Chandler::ToP2PMulShaUseGenCont::2012}. As an application-layer solution, P2P networks need to deal with many problems introduced with churn, i.e., the user-driven dynamics of peer participation. Most successful P2P networks, such as Napster, Gnutella, and PPLive, own millions of users. As such, churn in the networks may cause drastic changes in available bandwidth and end-to-end delay between peers. To maintain good performance, P2P networks need to dynamically adjust their (application-layer) network topology, i.e., a peer needs to dynamically find other ``good" peers for content delivery.

\begin{figure}[htb]
\begin{center}
\epsfig{file=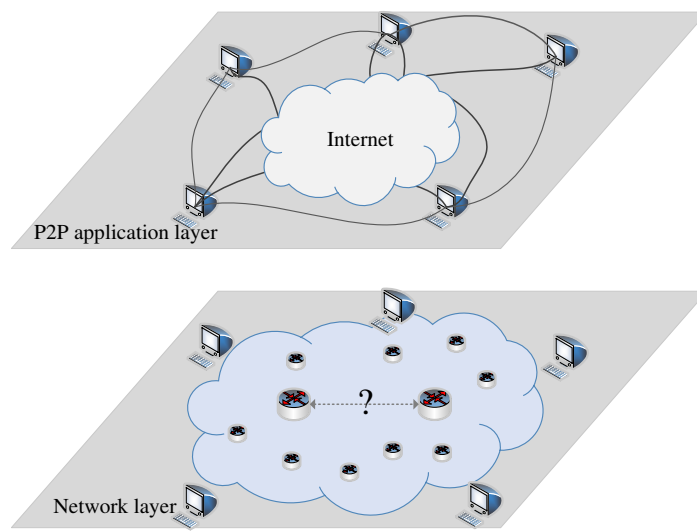, width=0.5\textwidth}
\vspace{-4mm}
\end{center}
\caption{\label{fig::P2PIP} A peer-to-peer network and the routing topology underneath}
\vspace{2mm}
\end{figure}

Nevertheless, peers normally do not have enough information to make a good decision in the path selection, due to the semantic gap between P2P networks and the physical networks underneath. In particular, the link between two neighboring peers in a P2P network is virtual in the sense that they usually do not have a direct physical link to connect them. Instead they rely on the underlying Internet to provide a communication path between them. As shown in Fig.~\ref{fig::P2PIP}, the P2P network (the top of the figure) depends on the physical network (the bottom of the figure) for packet delivery. Clearly, if peers are knowledgeable of the information regarding the routers and their connectivity, they can make much wiser decisions in peer selection. For example, a node may want to know the route topology to others so that it can choose peers with low or no route overlap to improve resilience against network failures~\cite{H.B::RON::2001}.

The task of discovering the network-layer topology under the hood of a P2P network can be fulfilled with two approaches. The first approach is based on the measurements or feedback messages of intermediate routers. This approach may not always work because some (anonymous) routers may not respond to probing messages (e.g., messages from \emph{traceroute})~\cite{R.V.B::TIAR::2003}. The second approach does not require the cooperation of intermediate routers, and it purely based on the end-to-end measurements controlled at end hosts. It utilizes the correlation among observed performance measure at end hosts to infer the internal network structure. This approach is termed as network tomography~\cite{Y.Var::NTESD::1996}.

Network tomography can be implemented in either an active or passive way. The active network tomography~\cite{Rab::MulSouMulDesNT::2004, CaceresR::MulInfNetInterLos::1999,FraLo::MulInfNetInterDis::2002} needs to explicitly send out probing messages to estimate the end-to-end path characteristics, while passive network tomography~\cite{JinCao::TimeVarNT::2000,Venkata::PassNTBayInf::2002,FabioRicciato::PasTom3GNet::2006,PasNTomErrNetNCAppr::HYao::2012} infers network topology without sending any explicit probing messages. In P2P networks, active network tomography may incur too large control overhead due to the sheer size of P2P networks and the frequent peer changes. Clear, passive network tomography is preferred and aligns very well with the context of P2P networks, because the data messages in P2P networks exhibit diverse (application-layer) forwarding paths and thus provide abundant end-to-end measures for network tomography.

We are thus motivated to design a passive network tomography that takes a free ride of content distribution in P2P networks with end-to-end delay correlation estimation (DCE). To infer the routing tree~\footnote{Rooted from a single source node, the network-layer routing structure to multiple destination nodes is always a tree structure at any given time instance, or otherwise routing loops exist.} rooted from a source node, the DCE method only needs the destination nodes to record the arriving times of messages from the source node. This requirement essentially puts zero control overhead for  network tomography and avoids the many problems in existing network tomography methods, such as Network Radar~\cite{YolanTsa::NetRadar::2004}, which use round-trip-time (RRT) measurement\footnote{We argue that passive network tomography using RRT measurement is problematic in P2P networks for at least two reasons: first, message delivery between peers may not use TCP, and thus we cannot use TCP-SYN and TCP-ACK for passive RRT estimation as in~\cite{YolanTsa::NetRadar::2004}; second, even if passive RRT estimation is possible, the correlation estimation of the return paths is hard due to the variable processing delay at the destination nodes.}.

The contributions of this paper include:

\begin{itemize}
  \item     We propose a new delay correlation estimation (DCE) methodology for estimating the characteristics of shared links from a source node to multiple destination nodes. Our method does not need special cooperation and synchronization between end hosts, making it largely different from prior tomography tools~\cite{YolanTsa::NetDelayTomo::2003} and \cite{YolanTsa::NetRadar::2004}. Moreover, it is simple to implement and scales well to large networks.
  \item  	We present a topology tomography algorithm more effective than the state-of-the-art Deepest-First-Search (DFS) ordering method introduced in~\cite{BDEriksson::ToPraNTIntTopoDiscov::2010}. Our algorithm is capable of handling dynamic scenarios where peers may join and leave the network frequently.
  \item 	We implement and test our passive network tomography approach over a small-scale Internet testbed as well as over PlanetLab. We also perform extensive simulation study in OMNeT++ to evaluate our method in large-scale networks. Both real-world experimental results and simulation results indicate that our method is accurate, robust, and scalable.
\end{itemize}

The rest of the paper is organized as follows. In Section~\ref{sec::delay correlation} we propose the method for delay correlation estimation (DCE). In Section~\ref{sec::topology tomography} we present topology recovery algorithm based on DCE. \nop{In Section~\ref{sec::passively realized} we propose a mechanism to passively realize topology tomography using network coded flow in P2P network.} In Sections~\ref{sec::experiments} and~\ref{simulation}, we present the real-world experimental results for small-scale networks and the simulation results for large-scale networks, respectively. In Section~\ref{sec::related work} we review related work. Section~\ref{sec::conclusions} concludes the paper.

\section{Delay Correlation Estimation}
\label{sec::delay correlation}

\begin{figure}[thb]
\begin{center}
\epsfig{file=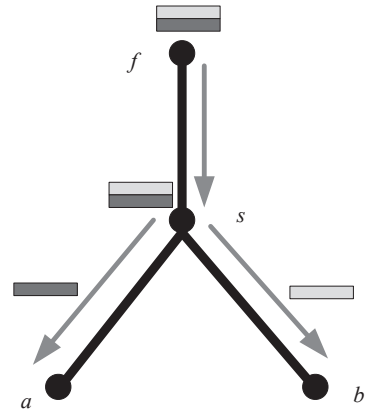, width=135pt}
\vspace{-4mm}
\end{center}
\caption{\label{fig::network_model} The tree structure: a sender \emph{f} and two receivers \emph{a, b}}
\vspace{-4mm}
\end{figure}

We first introduce the method for delay correlation estimation, which is the foundation of our new network tomography approach. To ease understanding, we use a simple example, which includes one sender $f$ and two receivers $a$ and $b$ as shown in Fig.~\ref{fig::network_model}, to illustrate the idea.

The routing structure from the sender $f$ to the receivers $a$ and $b$ must be a tree rooted at $f$. Otherwise, there is routing loop which must be corrected. Assume that router $s$ is the ancestor node of both $a$ and $b$. Assume that the sender uses unicast to send messages to receivers, and assume that packets are sent in a back-to-back pair\footnote{This implies that the sender have data to both receivers. This assumption is reasonable due to the diverse data forwarding requirements in P2P networks. Otherwise, passive network tomography will not have enough information to accurately infer routing topology.}. For the $k$-th pair of back-to-back packets, denoted as $a^k$ and $b^k$, sent from $f$ to $a$ and $b$, respectively, we use the following notation:
\begin{itemize}
	\item $t_a(k)$: the time when $a$ receive $a^k$ in the $k$-th pair.
	\item $t_b(k)$: the time when $b$ receive $b^k$ in the $k$-th pair.
	\item $d_a(k)$: the latency of $a^k$ along the path from $f$ to $a$.
	\item $d_a(k)$: the latency of $b^k$ along the path from $f$ to $b$.
  \item $t_f(k)$: the time when $f$ sends the $k$-th pair of packets.
\end{itemize}

Note that we can safely ignore the difference between the time when $f$ sends out $a^k$ and the time when it sends out $b^k$, because the two packets are back-to-back and the transmission time is negligible compared to the path latency. Therefore, for $a^0$ in the first pair of packets (we start the index with $0$ for convenience), we have
\begin{equation}
\label{eqn::t_a(0)=t_f(0)+d_a(0)}
t_a(0)=t_f(0)+d_a(0).
\end{equation}

Similarly, for $a^k$, we have
\begin{equation}
  \label{eqn::t_a(k)=t_f(k)+d_a(k)}
  t_a(k)=t_f(k)+d_a(k).
\end{equation}

Define $\delta_a(k) \equiv t_a(k)-t_a(0)$. We can obtain Eq.~(\ref{eqn::a(k)delta =(t_f(k)-t_f(0))+(d_a(k)-d_a(0))}) by letting Eq.(\ref{eqn::t_a(k)=t_f(k)+d_a(k)})$-$Eq.(\ref{eqn::t_a(0)=t_f(0)+d_a(0)}):

\begin{equation}
  \label{eqn::a(k)delta =(t_f(k)-t_f(0))+(d_a(k)-d_a(0))}
  \delta_a(k) =(t_f(k)-t_f(0))+(d_a(k)-d_a(0)).
\end{equation}

Denote the time interval between two consecutive pairs of packets as $\delta$. We assume that $\delta$ is a constant for simplicity at this moment, and later relax this assumption. If $\delta$ is constant, we can use $k\cdot\delta$ to replace $t_f(k)-t_f(0)$ in Eq.(\ref{eqn::a(k)delta =(t_f(k)-t_f(0))+(d_a(k)-d_a(0))}). We then have:
\begin{equation}
  \label{eqn::d_a(k)=a(k)delta-kdelta+d_a(0)}
  d_a(k)=\delta_a(k)-k\cdot\delta+d_a(0).
\end{equation}

Let $\delta'_a(k) \equiv \delta_a(k)-k\delta$. Eq.~(\ref{eqn::d_a(k)=a(k)delta-kdelta+d_a(0)}) can be rewritten as Eq.~(\ref{eqn::d_a(k)=a(k)'delta+d_a(0)})in the following.
\begin{equation}
  \label{eqn::d_a(k)=a(k)'delta+d_a(0)}
  d_a(k)=\delta'_a(k)+d_a(0).
\end{equation}

Similarly, we can obtain the result at receiver $b$ as in Eq.(\ref{eqn::d_b(k)=b(k)'delta+d_b(0)}):
\begin{equation}
  \label{eqn::d_b(k)=b(k)'delta+d_b(0)}
  d_b(k)=\delta'_b(k)+d_b(0),
\end{equation}
where $\delta'_b(k)\equiv \delta_b(k)-k\cdot \delta$, and $\delta_b(k) \equiv t_b(k)-t_b(0)$

To estimate the correlation between $d_a(k)$ and $d_b(k)$,  we introduce the following lemma.
\begin{Lemma}
  \label{Lemma::a,b,c,d}
  Assume that $\zeta$, $\eta$ are two random variables, and $\chi=a\zeta+b$, $\gamma=c\eta+d$, where $a,b,c,d$ are constants and $a,c$ have the same symbol. We have the correlation between $\zeta$ and $\eta$ is the same as the correlation between $\chi$ and $\gamma$, i.e.,  $\sigma_{\zeta,\eta}^{2}=\sigma_{\chi,\gamma}^{2}$.
\end{Lemma}

\begin{proof}
\begin{equation}
  \begin{split}
    \sigma_{\chi,\gamma}^{2} & =\frac{E(\chi-E\chi)(\gamma-E\gamma)}{\sqrt{D\chi}\sqrt{D\gamma}}\\
    & =\frac{E(a\zeta+b-aE\zeta-b)(c\eta+d-cE\eta-d)}{\sqrt{a^2D\zeta}\sqrt{c^2D\eta}}\\
    & =\frac{acE(\zeta-E\zeta)(\eta-E\eta)}{|a||c|\sqrt{D\zeta}\sqrt{D\eta}}\\
    & =\frac{E(\zeta-E\zeta)(\eta-E\eta)}{\sqrt{D\zeta}\sqrt{D\eta}}\\
    & =\sigma_{\zeta,\eta}^{2}
  \end{split}
  \end{equation}
\end{proof}

Note that in any continual serial of $n$ packets, both $d_a(0)$ and $d_b(0)$ can be regarded as constants. Thus, based on Lemma \ref{Lemma::a,b,c,d}, we have the following theorem:
\begin{Theorem}
  \label{Therorem::correlation between a delta and b delta}
  The correlation between delay variables $d_a(k)$ and $d_b(k)$ is equal to the correlation between variables $\delta'_a(k)$ and $\delta'_b(k)$, that is,
  \begin{equation}
    \sigma_{d_a(k),d_b(k)}^{2}=\sigma_{\delta'_a(k),\delta'_b(k)}^{2}.
  \end{equation}
\end{Theorem}

Based on the measurements of $\delta'_a(k)$, $\delta'_b(k)$, we can calculate the correlation of delays along the path from $f$ to $a$ and along the path from $f$ to $b$, denoted as $\sigma_{\delta'_a,\delta'_b}^{2}$, using Eq.(\ref{equ::a'delta,b'delta}):

\begin{equation}
  \label{equ::a'delta,b'delta}
  \sigma_{\delta'_a,\delta'_b}^{2}=\frac{1}{n-1}\sum_{k=1}^{n}[\delta'_a(k)-\overline{\delta'_a}][\delta'_b(k)-\overline{\delta'_b}]
\end{equation}
where $n$ is the total number of measurements used for estimation, $\overline{\delta'_a}$ is the sample mean of ${\delta'_a(k)}_{k=1}^{n}$, and $\overline{\delta'_b}$ is the sample mean of ${\delta'_b(k)}_{k=1}^{n}$. Following a similar argument as in~\cite{YolanTsa::NetRadar::2004}, it is easy to conclude:
\begin{Theorem}
  \label{Therorem::correlation unbiased estimator}
  Eq.(\ref{equ::a'delta,b'delta}) is an unbiased estimator of $\sigma_{\delta'_a,\delta'_b}^{2}$.
\end{Theorem}

The pseudocode of DCE is summarized in Algorithm~\ref{alg:DCEBased on PATI}.

\begin{algorithm}
\caption{Delay Correlation Estimation}
\label{alg:DCEBased on PATI}
\begin{algorithmic}[1]
 \REQUIRE ~~\\
 Given the time interval $\delta$ for sending packet pairs, number of packet pairs $n$.
 \ENSURE~~\\
 \FOR {$k=0:n$}
    \STATE Use Eq.~(\ref{eqn::t_a(0)=t_f(0)+d_a(0)}) to Eq.~(\ref{eqn::a(k)delta =(t_f(k)-t_f(0))+(d_a(k)-d_a(0))}) to measure $\delta_a(k)$ in the $k$-th transmission;
    \STATE Use $\delta'_a(k)=\delta_a(k)-k\delta$ to calculate $\delta'_a(k)$ in the $k$-th loop;
    \STATE Similarly, measure $\delta_b(k)$ and calculate $\delta'_b(k)=\delta_b(k)-k\delta$;
 \ENDFOR
 \STATE With all the values of $\delta'_a(k)$ and $\delta'_b(k)$, use Eq.(\ref{equ::a'delta,b'delta}) to obtain the correlation $\sigma_{\delta'_a,\delta'_b}^{2}$, which is equivalent to the delay correlation $\sigma_{a,b}^2\equiv\sigma_{d_a,d_b}^{2}$ between hosts $a$ and $b$ according to Theorem~\ref{Therorem::correlation between a delta and b delta}.
\end{algorithmic}
\end{algorithm}

\begin{remark}
In Algorithm~\ref{alg:DCEBased on PATI}, we assumed that the time interval $\delta$ for sending packet pairs is constant. With a slight change, Algorithm~\ref{alg:DCEBased on PATI} works well when $\delta$ is not constant. In this case, $t_f(k)-t_f(0)\neq k\cdot \delta$, and thus using $k\cdot \delta$ to replace $t_f(k)-t_f(0)$ is inappropriate. In this case we choose $\delta_f(k)$ to denote $t_f(k)-t_f(0)$ in Eq.~(\ref{eqn::a(k)delta =(t_f(k)-t_f(0))+(d_a(k)-d_a(0))}), then Eq.~(\ref{eqn::d_a(k)=a(k)delta-kdelta+d_a(0)}) can be re-written to Eq.(\ref{equ::d_a(k)=delta_a(k)-delta_f(k)+d_a(0)}):
\begin{equation}
  \label{equ::d_a(k)=delta_a(k)-delta_f(k)+d_a(0)}
  d_a(k)=\delta_a(k)-\delta_f(k)+d_a(0).
\end{equation}
Correspondingly, Eq.~(\ref{eqn::d_a(k)=a(k)'delta+d_a(0)}) and Eq.~(\ref{eqn::d_b(k)=b(k)'delta+d_b(0)}) can be replaced with $\delta'_a(k)=\delta_a(k)-\delta_f(k)$ and $\delta'_b(k)=\delta_b(k)-\delta_f(k)$, respectively. Note that each $t_f(k)$ is a timestamp contained in the packet, and thus $\delta_f(k)=t_f(k)-t_f(0)$ is readily available.
\end{remark}

\section{Dynamic Topology Tomography with DCE}
\label{sec::topology tomography}

After obtaining the DCE values, we now use them to infer (network-layer) routing topology. We assume that the routing tables do not change during a short time period, and thus there is one unique active path between every two end hosts during the measurement period. Therefore, if the source sends packets to a set of destination hosts, the paths from the source node to the destination nodes form a tree structure determined by the underlying routing algorithm. The tree structure is the common assumption for topology inference starting with a \textit{single} source. Otherwise, current router configuration is problematic: either there is a routing loop in the network routing tables, or some routes could be optimized to become shorter.

\begin{figure*}[htb]
\begin{center}
\begin{minipage}{40mm}
\centering
\epsfig{file=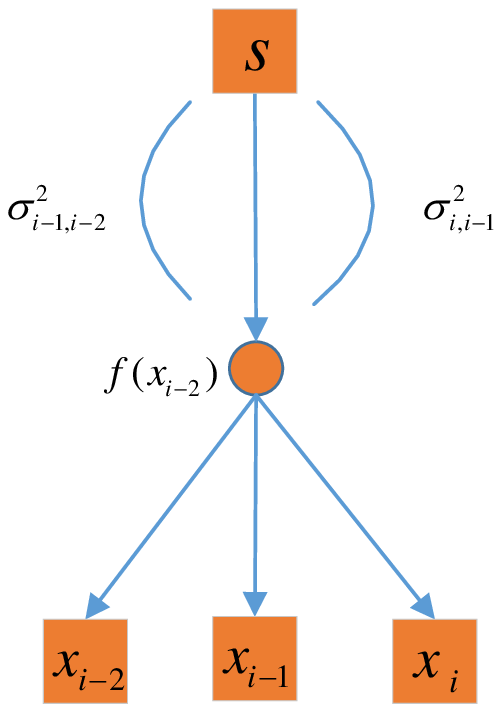, width=102pt}
\vspace{-2mm}
\caption{\label{fig::case1} $|\sigma_{i-1,i-2}^2-\sigma_{i,i-1}^2|<\varrho$.}
\vspace{-1mm}
\end{minipage} \hfil
\begin{minipage}{40mm}
\centering
\epsfig{file=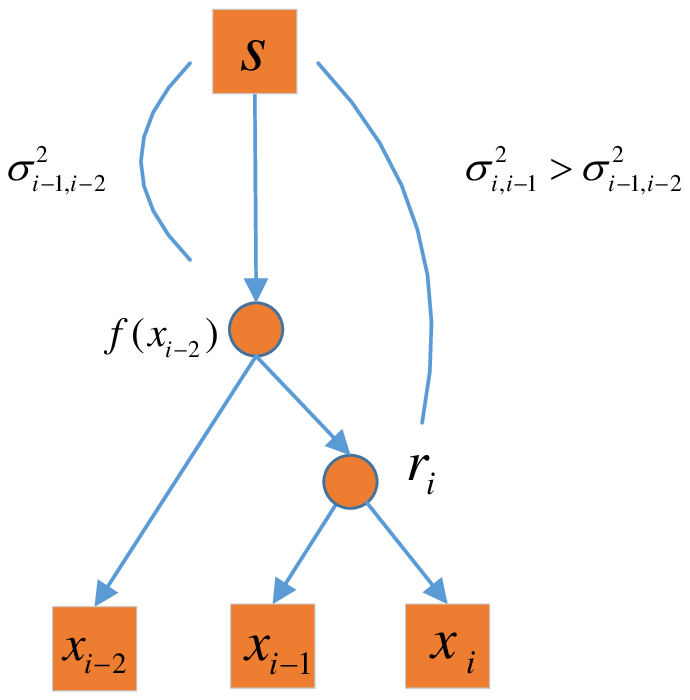, width=145pt}
\vspace{-7mm}
\caption{\label{fig::case2} $\sigma_{i,i-1}^2\geq\sigma_{i-1,i-2}^2+\varrho$.}
\vspace{-2mm}
\end{minipage} \hfil
\begin{minipage}{40mm}
\centering
\epsfig{file=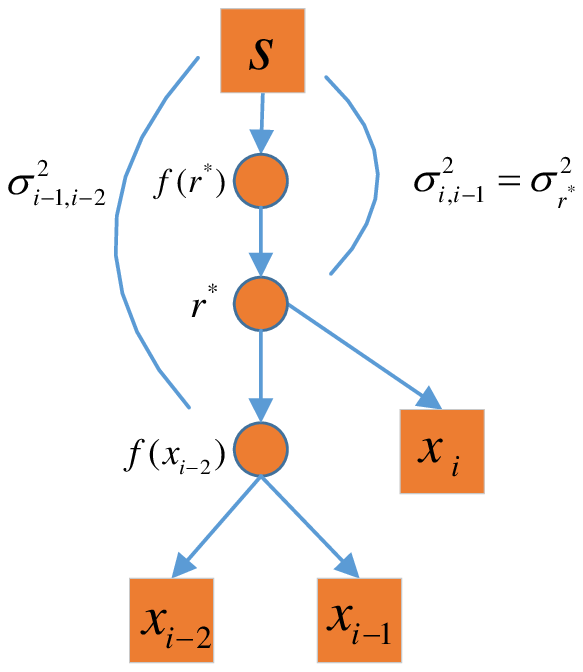, width=140pt}
\vspace{-7mm}
\caption{\label{fig::case3} $|\sigma_{r^{\star}}^2-\sigma_{i,i-1}^2|<\varrho$.}
\vspace{-1mm}
\end{minipage} \hfil
\begin{minipage}{40mm}
\centering
\epsfig{file=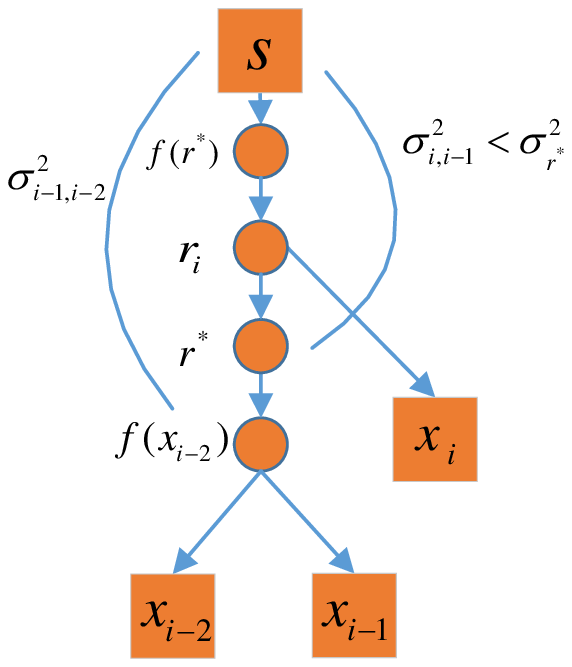, width=140pt}
\vspace{-7mm}
\caption{\label{fig::case4} $\sigma_{r^{\star}}^2\geq\sigma_{i,i-1}^2+\varrho$.}
\vspace{-1mm}
\end{minipage} \hfil
\begin{minipage}{40mm}
\centering
\epsfig{file=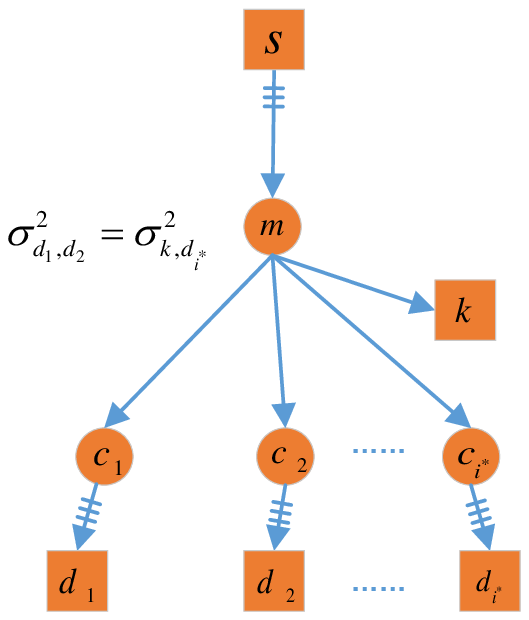, width=115pt}
\vspace{-2mm}
\caption{\label{fig::caseadd1} $|\sigma_{d_1,d_2}^2-\sigma_{k,d_{i^\star}}^2|<\varrho$.}
\vspace{-7mm}
\end{minipage} \hfil
\begin{minipage}{40mm}
\centering
\epsfig{file=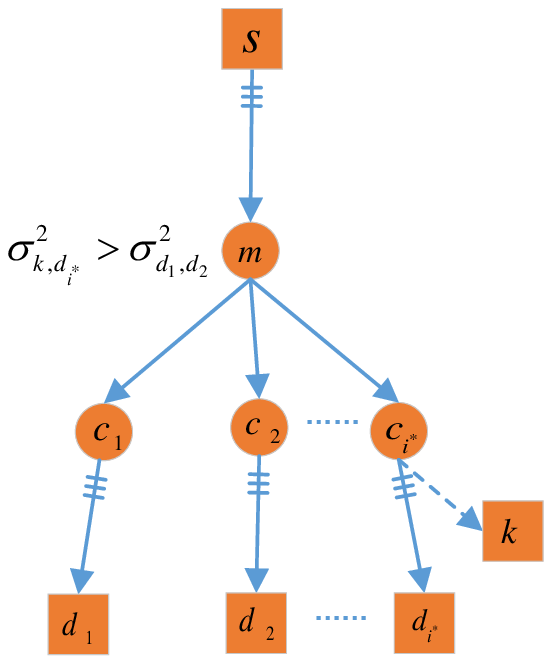, width=117pt}
\vspace{-2mm}
\caption{\label{fig::caseadd2} $\sigma_{k,d_{i^\star}}^2-\sigma_{d_1,d_2}^2\geq\varrho$.}
\vspace{-7mm}
\end{minipage} \hfil
\begin{minipage}{40mm}
\centering
\epsfig{file=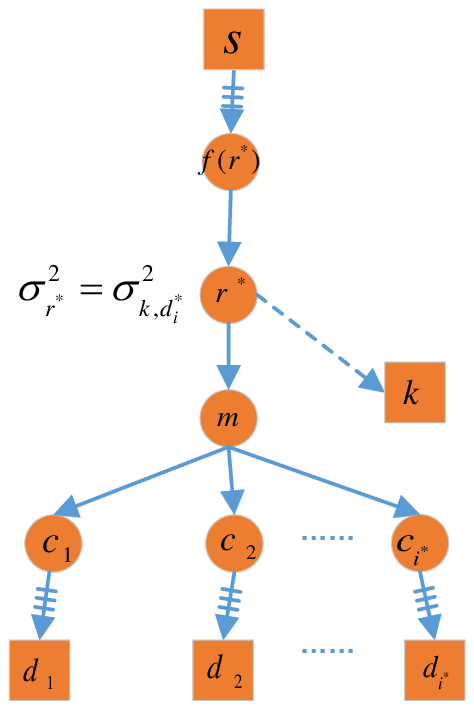, width=100pt}
\vspace{-2mm}
\caption{\label{fig::caseadd3} $|\sigma_{r^\star}^2-\sigma_{k,d_{i^\star}}^2|<\varrho$.}
\vspace{-7mm}
\end{minipage} \hfil
\begin{minipage}{40mm}
\centering
\epsfig{file=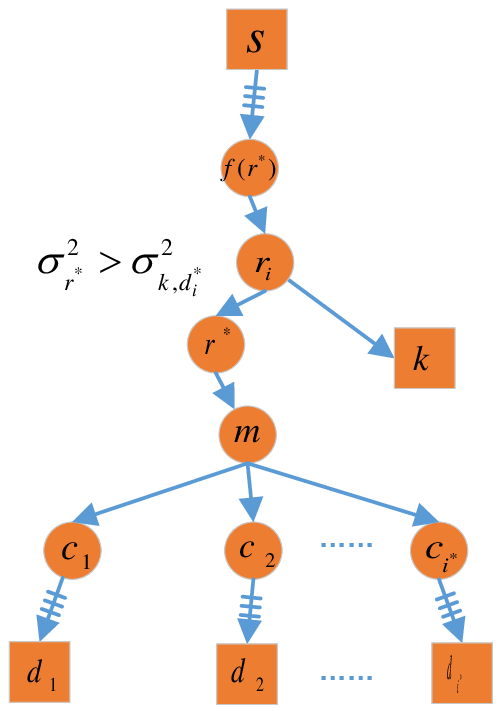, width=105pt}
\vspace{-2mm}
\caption{\label{fig::caseadd4} $\sigma_{r^\star}^2\geq\sigma_{k,d_{i^\star}}^2+\varrho$.}
\vspace{-7mm}
\end{minipage} \hfil
\end{center}
\end{figure*}

It has been shown in~\cite{BDEriksson::ToPraNTIntTopoDiscov::2010} that if the Depth-First-Search (DFS) order of the leaf nodes of a tree is given, i.e., the leaf nodes are arranged in the same order as that obtained by the DFS algorithm, then we can effectively recover the tree topology. Nevertheless, the DFS order of the leaf nodes are not directly available in practice. To solve this problem, a bisection DFS ordering algorithm, called \textit{bisect}, is proposed in~\cite{BDEriksson::ToPraNTIntTopoDiscov::2010} to estimate the DFS order of destination hosts. Although the paper~\cite{BDEriksson::ToPraNTIntTopoDiscov::2010} uses round-trip-time (RRT) values, we can also use \textit{bisect} to obtain the DFS order of leaf nodes using DCE instead of RRT values. This is because if we ignore the processing time at the destination nodes, the delay correlation obtained with RRT and the delay correlation obtained with our one-way measures are essentially the same. Since the \textit{bisect} algorithm is not key to understand our method, we omit its detail and in the rest of the paper assume that a set of leaf nodes, given their pairwise DCE values w.r.t the same root node (source), could be arranged in the DFS order.

After we obtain the DFS order of leaf nodes, we can follow the tree recovery algorithm in~\cite{BDEriksson::ToPraNTIntTopoDiscov::2010} to build the tree topology. Furthermore, we need to enhance the algorithm to quickly re-construct the tree without re-run the whole process again, when a peer joins/leaves the network. For this paper to be self-contained, we first use examples to reiterate the tree recovery algorithm in~\cite{BDEriksson::ToPraNTIntTopoDiscov::2010}, which is referred as static network tomography in this paper.

%


%
\subsection{Static Network Tomography}

We illustrate how to construct a tree topology when a set of hosts $\langle x_1', x_2',...,x_N'\rangle$ receives messages from a common peer \emph{s}. In this case, we can recover a tree rooted at \emph{s} and with $x_1', x_2',...,x_N'$ as leaf nodes. The intermediate nodes of the tree are routers. As stated above, we assume that the leaf nodes have been arranged in the DFS order, using the DCE values and the \textit{bisect} algorithm in~\cite{BDEriksson::ToPraNTIntTopoDiscov::2010}. Assume that $\varrho$ is the minimum possible delay covariance induced by a single router. Topology structure could be recovered in the following different cases shown in Fig.~\ref{fig::case1}-Fig.~\ref{fig::case4}.

\begin{itemize}
	\item \textbf{Case 1}: $|\sigma_{i-1,i-2}^2-\sigma_{i,i-1}^2|<\varrho$. This condition implies that the shared path between node pair $\langle x_{i},x_{i-1}\rangle$ and the shared path between node pair $\langle x_{i-1},x_{i-2}\rangle$ consist of exactly the same set of routers, as shown in Fig.~\ref{fig::case1}. Note that $f(x_{i-2})$ means the father node of $x_{i-2}$.
	\item \textbf{Case 2}: $\sigma_{i,i-1}^2\geq\sigma_{i-1,i-2}^2+\varrho$. This condition implies the shared path between node pair $\langle x_i,x_{i-1}\rangle$ is longer than the shared path between node pair $\langle x_{i-1},x_{i-2}\rangle$. In this case we create a new router $r_i$ with $x_i,x_{i-1}$ as its children and $f(x_{i-2})$ as its father node, as shown in Fig.~\ref{fig::case2}. For each added router $r_i$, the covariance value associated with the shared path to $r_i$ needs to be recorded. For this purpose, we denote $\sigma^2_{r_i}=\sigma^2_{i,i-1}$.
\item \textbf{Case 3}: $\sigma_{i,i-1}^2+\varrho<\sigma_{i-1,i-2}^2$. This condition implies that $x_i$ should be attached to a router higher in the tree than $f(x_{i-1})$. In this case we first find the farthest router $r^{\star}$  with $\sigma_{r^{\star}}^2\geq\sigma_{i,i-1}^2$ from current node $x_{i-1}$.
\begin{itemize}
 \item \textbf{Case 3.1}: $|\sigma_{r^{\star}}^2-\sigma_{i,i-1}^2|<\varrho$. This case is similar as the case in Fig.~\ref{fig::case1}, and we attach $x_i$ to $r^{\star}$ directly, as shown in Fig.~\ref{fig::case3}.
\item \textbf{Case 3.2}: $\sigma_{r^{\star}}^2\geq\sigma_{i,i-1}^2+\varrho$. This condition implies there is a hidden router not found between $r^{\star}$ and its parent $f(r^{\star})$. We create a router $r_i$ between them and attach $x_i$ to it, as shown in Fig.~\ref{fig::case4}. We also need to record $\sigma^2_{r_i}=\sigma^2_{i,i-1}$.
\end{itemize}
\end{itemize}

\subsection{Dynamic Network Tomography}

In a P2P network, peers may join and leave frequently. It is thus necessary to learn new network-layer routing changes, particularly when new peers join and potentially select routers which have not been discovered before. For this purpose, we should not re-run the above process from scratch, but instead should infer routing topology based on existing known structure. We call this procedure dynamic network tomography. We consider another four cases when a peer $k$ joins the P2P network. Given the existing tree, assume that $k$ is attached to a router $m$, called \emph{base router}, which is initially unknown. Our task is to determine the location of $m$. Initially, we select an intermediate node\footnote{For simplicity we usually choose the root node as the base router. However, in fact any intermediate node can be selected as the base router since the dynamic algorithm can go both directions in different cases within a tree.} in the tree as the base router and then adjust its position or create a new router if no adjustment is possible.

Assume that the current base router $m$ has $l$ children, denoted as $c_1,c_2,\ldots,c_l$, respectively. For each $c_i (i=1, 2, \ldots, l)$, we select a destination node (i.e., leaf node) $d_i$ descended from $c_i$. Note that $d_i=c_i$ if $c_i$ is a leaf node already. Select any pair of nodes from $l$ leaf nodes, $d_1, d_2, \ldots, d_l$, say $d_1, d_2$ without losing generality. Since a pair of leaf nodes only shares the path from the root $s$ to $m$, the delay correlation values of any two pair of leaf nodes should be similar (i.e., bounded by $\varrho$). We utilize Algorithm~\ref{alg:DCEBased on PATI} to calculate $\sigma_{d_1, d_2}^2$ and $\sigma_{k,d_i}^2$ for $i=1,...,l$. We then find the leaf node, denoted as $d_{i^\star}$, that has the largest delay correlation with $k$. Correspondingly, in the tree branch where $d_{i^\star}$ sits, we denote the direct child of $m$ as $c_{i^\star}$.

The following different cases shown in Fig.~\ref{fig::caseadd1}-Fig.~\ref{fig::caseadd4} require different ways to identify the location of the base router.

\begin{itemize}
	\item \textbf{Case 5}: $|\sigma_{d_1,d_2}^2-\sigma_{k,d_{i^\star}}^2|<\varrho$. This condition implies that the shared path between node pairs $\langle d_1,d_2\rangle$ and the shared path between node pair $\langle k,d_{i^\star}\rangle$ are the same. Therefore, $k$ is a child of $m$ and should be attached to it, as shown in Fig.~\ref{fig::caseadd1}.
	\item \textbf{Case 6}: $\sigma_{k,d_{i^\star}}^2-\sigma_{d_1,d_2}^2\geq\varrho$. This condition implies the shared path between node pair $\langle k,d_{i^\star}\rangle$ is longer than the shared path between node pair $\langle d_1,d_2\rangle$. In this case, $k$ should belong to the tree branch rooted from $c_{i^\star}$. We then adjust $c_{i^\star}$ as the new base router, and perform the same procedure (i.e., check Cases 5, 6, 7) again.
	\item \textbf{Case 7}: $\sigma_{k,d_{i^\star}}^2+\varrho<\sigma_{d_1,d_2}^2$. This implies that $k$ should be attached to a router higher in the tree than $m$. In this case we first find the farthest router $r^{\star}$  with $\sigma_{r^{\star}}^2 \geq \sigma_{k,d_{i^\star}}^2$.	
\begin{itemize}
	\item \textbf{Case 7.1}: $|\sigma_{r^\star}^2-\sigma_{k,d_{i^\star}}^2|<\varrho$. In this case, let $k$ join via node $r^\star$, as shown in Fig.~\ref{fig::caseadd3}.
	\item \textbf{Case 7.2}: $\sigma_{r^\star}^2\geq\sigma_{k,d_{i^\star}}^2+\varrho$. This condition implies that there is a hidden router not found between router $r^\star$ and its parent $f(r^{\star})$. We create a new router $r_i$ between them to attach $k$ to it, as shown in Fig.~\ref{fig::caseadd4}. We also record $\sigma^2_{r_i}=\sigma_{k,d_{i^\star}}^2$
\end{itemize}
\end{itemize}

It is easy to handle the case when a peer leaves. We just simply remove it and the associated edges from the tree.

\nop{
A big difference from \cite{BDEriksson::ToPraNTIntTopoDiscov::2010} is that we consider network dynamics. To add a new node in the tree the overhead of total packet needed is only $l\log_lN$ for DCE compared to $\frac{N}{2}l\log_lN$ for DFS. The complexity of this methodology is reduced by one order of magnitude $O(N)$. Therefore, our approach is capable of both static and dynamic scenarios especially when size of end hosts is large.

Since it is able to trace bidirections in the tree if we choose an appropriate intermediate node \emph{m} (Take the center node for example) the total overhead can be further reduced from $l\log_lN$ to $\frac{1}{2}l\log_lN$.

In addition, DCE based tomography only needs single trip time measurement to calculate delay correlations compared to the RTT approach in \cite{BDEriksson::ToPraNTIntTopoDiscov::2010}, which saves 50\% time cost, thus is time efficient. Moreover, it is valid in theory while RTT based approach violates an essential assumption that the return paths are uncorrelated but a segment of the return path (See network model in \emph{Fig.\ref{fig::network_model}}) is shared actually. }

\section{Experiments over Real Networks}
\label{sec::experiments}
We implemented our passive network tomography approach over real world networks, including a small-scale Internet platform and PlanetLab. As a comparison, we also implemented the DFS ordering method~\cite{BDEriksson::ToPraNTIntTopoDiscov::2010} and the sequential method~\cite{JianNi::EffDynaRouTopoInfEnd2End::2010}.

The DFS ordering method works in a similarly way like our DCE topology recovery method, and the sequential network tomography method constructs network topology by adding each host sequentially, i.e., step by step. All the methods, including ours, use the same framework for network tomography: For a host \emph{h} and an intermediate node \emph{c}, the correlations between \emph{h} and the descendants of \emph{c} are found to determine the location of \emph{h}.

The main difference between our method and the DFS ordering method is that the latter utilizes $RTT$ values and cannot handle dynamic scenario when end hosts join/leave frequently. The sequential method reduces the computational complexity for tree construction, but when the largest correlation value between \emph{h} and the descendants of \emph{c} is smaller than the variance of node \emph{c}, it may not find the right location of \emph{h}. In addition, the experimental study in~\cite{JianNi::EffDynaRouTopoInfEnd2End::2010} utilizes one way delay,  which is not easy to measure precisely in practice due to the time synchronization issues among end hosts. In contrast, our method does not need strict time synchronization, since it only needs to identify the pair of two back-to-back packets and record their arrival times at the destination nodes.

In order to determine the accuracy of recovered topology we choose a metric called tomography accuracy \emph{p}, which is also used in~\cite{BDEriksson::ToPraNTIntTopoDiscov::2010}.

\begin{equation}
  \label{eqn::frac{1}{|X|^3}}
  p=\frac{1}{|X|^3}\sum_{i\in{X}}\sum_{j\in{X}}\sum_{k\in{X}}f(i,j,k),
\end{equation}
where \emph{X} is the whole set of destination nodes, $f(i,j,k)=1$ if the reconstructed topology correctly classifies the triple $\{i,j,k\}$ and $f(i,j,k)=0$ otherwise. The classification rule is defined as follows:

\begin{equation}
  \begin{split}
     f(i,j,k) &=\varphi(\widehat{P}(i,j)\geq\widehat{P}(i,k))\varphi(P(i,j)\geq P(i,k))\\
     &+\varphi(\widehat{P}(i,j)<\widehat{P}(i,k))\varphi(P(i,j)<P(i,k)),
  \end{split}
\end{equation}
where $\widehat{P}(i,j)$ denotes the length of the shared path between end nodes $i,j$ on the recovered topology, $P(i,j)$ denotes the length of the shared path between end nodes $i,j$ on the \emph{ground truth topology}, $\varphi(x)=1$ if the condition $x$ holds and $\varphi(x)=0$ otherwise.

\subsection{Experiment over the Internet}

\begin{figure}[htb]
\begin{center}
\epsfig{file =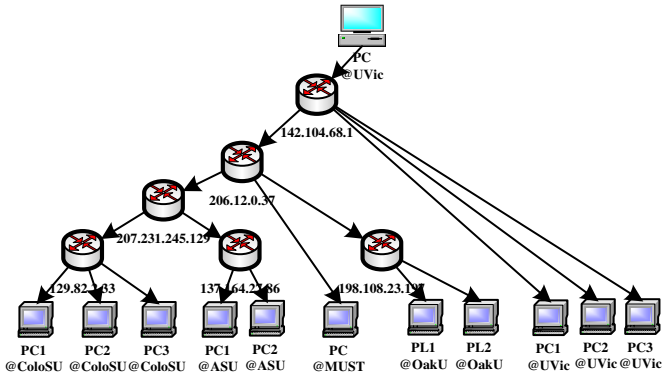, width=0.5\textwidth}
\vspace{-2mm}
\caption{\label{fig::not_planetlab_tree_topology} Topology of our Internet testbed}
\end{center}
\vspace{-4mm}
\end{figure}

For real world experiment we first deploy DCE tomography on a small-scale Internet platform with $12$ hosts locating in $5$ universities (shown in Fig.~\ref{fig::not_planetlab_tree_topology}). We select BitTorrent~\cite{BitTorrent::WebSite}, an open source project supporting peer-to-peer file sharing, as the P2P application. We implement a module, called \emph{tomo-component}, for data analysis. As regular data flows are sent from a source to destinations, \emph{tomo-component} utilizes the transmitted data for passive network tomography. There is a need to set the time interval $\delta$ between two pairs of back-to-back packets. We estimate the value of $\delta$ by using \textit{traceroute} probes. In this case study, the average one way delay on the longest path is about $30$ ms. Note that a too large $\delta$ value do not hurt network tomography but would unnecessarily slow down the data transmission rate. On the other hand, if $\delta$ is too small (e.g., $1000 \mu$s), intermediate routers may drop packets due to congestion. To evaluate the accuracy of our passive network tomography, we use \textit{traceroute} probes to obtain the \textit{ground truth routing topology}.

\begin{figure}[htb]
\begin{center}
\epsfig{file =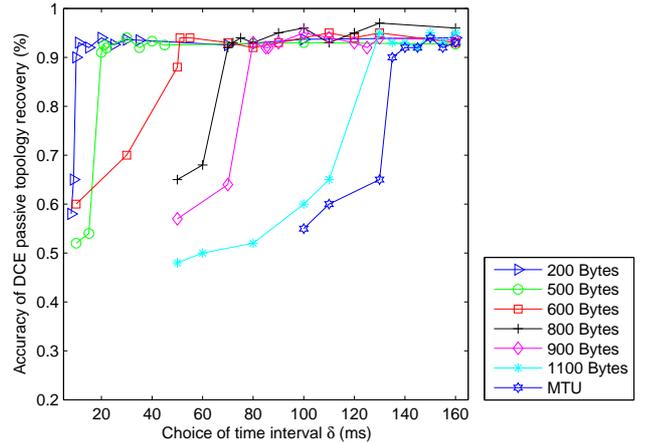,width=0.53\textwidth}
\vspace{-4mm}
\caption{\label{fig::accuracy_interval_packsize} DCE tomography accuracy influenced by $\delta$ (Internet experiment)}
\end{center}
\vspace{-4mm}
\end{figure}

\nop{The CPU usage of all 12 physical hosts is normal (Below 20\% most of time with no virtual machine running) and they are regarded as a graph network connecting by intermediate routers. }

In our test, we use a host in University of Victoria (UVic) as the source and the other hosts as the destination nodes of content distribution. Since there is no routing loop, data flows from the source to the destinations form a spanning tree, rooted at the host in UVic and with other hosts as leaf nodes (see Fig.~\ref{fig::not_planetlab_tree_topology} for details). We test the tomography accuracy with different size of data packets and different interval time of sending back-to-back packets. The results are shown in Fig.~\ref{fig::accuracy_interval_packsize}. From the figure we can see that when the packet size is small (e.g., $200$ bytes), our approach can achieve $92\%$ of accuracy in toplogy recovery even when the interval time is small. It is worth noting from the results we can see that passive network tomography does not work very well when the packet size is large and the interval time is small. This is easy to understand, since in this case the network may be congested and drops packets. As a result, DCE values may not capture the path correlation well.

\begin{figure}[htb]
\begin{center}
\epsfig{file =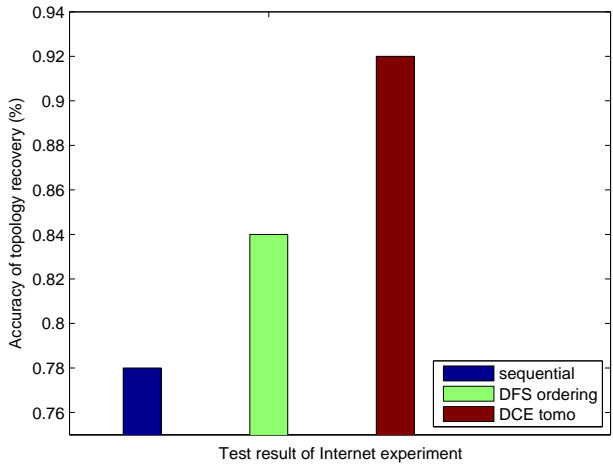,width=0.5\textwidth}
\vspace{-4mm}
\caption{\label{fig::result_on_internet} Performance comparison over the Internet testbed}
\end{center}
\vspace{-4mm}
\end{figure}

The size of packages should be carefully chosen to meet the requirement for regular data transmission and passive topology tomography. From the experimental results shown in Fig.\ref{fig::accuracy_interval_packsize}, when the interval time $\delta$ is small (e.g., $20$ ms), the max package size should also be small (e.g., $500$ bytes correspondingly). If the interval time $\delta$ is large enough (e.g., $135$ ms), the max package size can be large. Nevertheless, to avoid delay for package segmentation, we should not set the size of packets larger than the Maximum Transmission Unit (MTU) in practice.

For comparison purpose, we also implement the DFS ordering method using RTT and the sequential method using one way delay time. Fig.~\ref{fig::result_on_internet} shows the best performance each method can achieve in our test bed. We see clearly that the DCE tomography has the best performance of $92\%$ recovery accuracy on this Internet test bed. The processing delay at the destination nodes for the RTT measures and the inaccuracy in one-way delay measures may contribute to the worse performance of the DFS ordering method and the sequential method.

\subsection{Experiment Over PlanetLab}

\begin{figure*}[htb]
\begin{center}
\epsfig{file =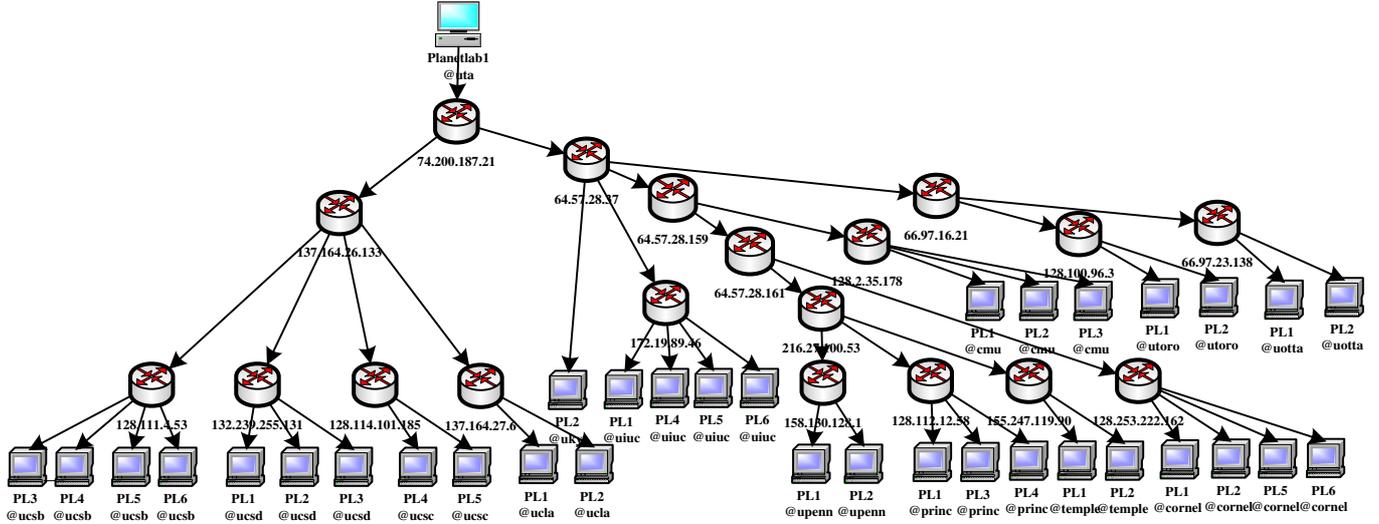,width=1\textwidth}
\vspace{-4mm}
\caption{\label{fig::tree_topology_on_PlanetLab} Topology of our PlanetLab testbed}
\end{center}
\vspace{-4mm}
\end{figure*}

PlanetLab is an openly available network testbed designed to support many different types of network-related experiments. It is comprised of over $1170$ end hosts deployed at about $550$ different sites all over the world~\cite{PlanetLab::WebSite}. We choose PlanetLab for two reasons: we can configure a larger real-world testbed with a manageable cost, and PlanetLab has its own special features which may impact network tomography. For this test, we choose $35$ nodes locating in North America on PlanetLab (Fig.~\ref{fig::tree_topology_on_PlanetLab}). We have to ignore PlanetLab nodes that cannot be found using \textit{traceroute}, since there are some routers not responding to \textit{traceroute} probes.

With the setting of package size of $500$ bytes and the interval time of $30$ ms, we can achieve over $90\%$ accuracy in the Internet testbed. With the same setting, however, our method over the above PlanetLab testbed can only achieve less than $40\%$ accuracy. There is a sharp performance drop over the PlanetLab platform.

Three main reasons account for the poor result:

\begin{enumerate}
	\item We observe that most PlanetLab nodes are always heavily loaded with multiple applications and sometimes their CPU utilization is nearly $100\%$! Hence, the workload on host systems can have a large impact on OS scheduling and significantly alter the times in content streaming. In particularly, the back-to-back property of packet pairs may be destroyed. Some measurement~\cite{KoungSoo::CoMon::2005} even points out that heavily loaded hosts are quite normal in PlatnetLab.
	\item PlanetLab uses virtualization techniques to isolate slice\footnote{A slice is a set of allocated resources distributed across PlanetLab. Slice is
implemented using a technique called distributed virtualization. After nodes have been assigned to a slice, virtual servers for that slice are created on each of the assigned nodes.} between users~\cite{SteSol::ContOpeSysVir::2007},~\cite{AndyMic::OpSySuPlanetNetSer::2004}; Resource scheduling in a virtualization environment may result in large variations in the DCE values.
	\item Large variations on the background traffic crossed the PlanetLab nodes~\cite{AcMeaSysSharEnv::JoelPaul::2007} may create a lot of noise in the DEC values.
\end{enumerate}

\begin{figure}[htb]
\begin{center}
\epsfig{file =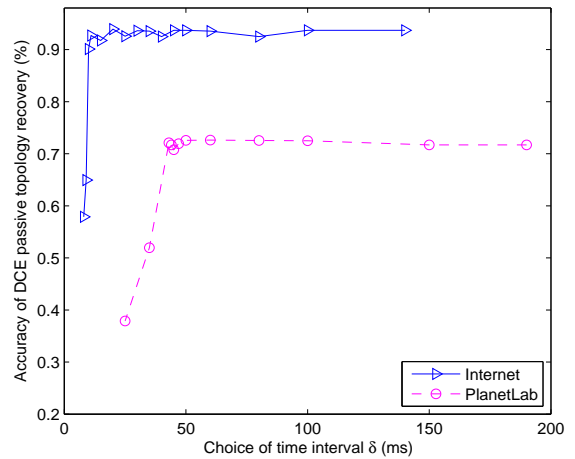,width=0.45\textwidth}
\vspace{-4mm}
\caption{\label{fig::accuracy_influence_delta_high_load} Comparison of DCE tomography accuracy influenced by $\delta$ on the Internet testbed and the PlanetLab testbed}
\end{center}
\vspace{-4mm}
\end{figure}

The above problems can be (slightly) alleviated by reducing the size of packets and increasing the time interval $\delta$ between consecutive pairs of back-to-back packets. When we fix the size of packets to $200$ bytes and change the time interval $\delta$ from $30$ ms to $50$ ms, we can observe an improvement on tomography accuracy from less than $40\%$ to about $70\%$. The results are shown in Fig.~\ref{fig::accuracy_influence_delta_high_load}. Nevertheless, further increasing the time interval $\delta$ does not lead to further performance improvement.

\begin{figure}[htb]
\begin{center}
\epsfig{file =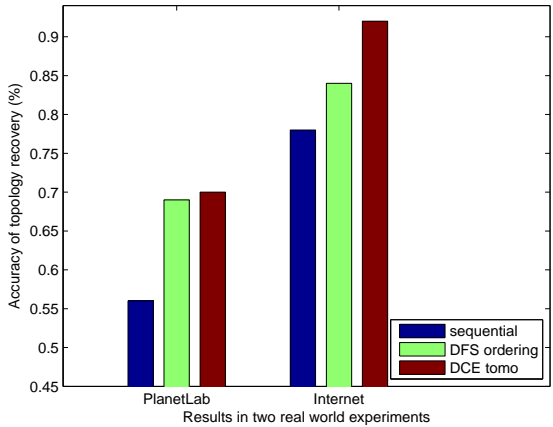,width=0.45\textwidth}
\vspace{-4mm}
\caption{\label{fig::result_in_two_experiment} Performance comparison over the Internet testbed and the PlanetLab testbed}
\end{center}
\vspace{-4mm}
\end{figure}

Similar with the experiment on the Internet testbed (Fig.~\ref{fig::not_planetlab_tree_topology}), we compare DCE based tomography with the DFS ordering method and the sequential method on the PlanetLab testbed. Fig.~\ref{fig::result_in_two_experiment} shows the best results that the three methods can achieve on the two testbeds. We find that although DCE tomography has the best recovery accuracy of 70\% on PlanetLab, there is still a large gap between its performance on the PlanetLab testbed and that on the Internet testbed. Our lessons learned from the PlanetLab testbed suggest that PlanetLab may not be a proper environment to test network tomography. It is hard to avoid the problems caused by high workload of end hosts and the virtualization based design of PlanetLab.

\section{Large-scale Simulation Evaluation}
\label{simulation}
\subsection{Setups of simulation}

We use simulation to further evaluate the performance of DCE tomography in a much larger network. The simulation environment is OMNeT++~\cite{OMNet++::WebSite}, which is an open-architecture discrete-event simulator consisting of extensible, modular, and component-based C++ simulation libraries. We use the tool of a representative topology generator BRITE to produce AS-level as well as router-level topology with different traffic patterns that can be used directly by OMNeT++.

We generate a random network with $200$ nodes with $150$ end hosts and $50$ intermediate routers. We choose the \emph{Waxman}\footnote{Waxman refers to a generation model for a random topology using Waxman's probability model for interconnecting the nodes.}~\cite{Amedina::Brite::2001} model to determine the connectivity between routers. Each time a host is selected randomly as the source and a subset ($70\%$) of others as clients. When the source transmits data to the clients, the routing algorithm determines a routing tree, rooted at the source and with the clients as the leaf nodes. We set the bandwidth of each link as $100$ Mbps. The rest subset of $30\%$ hosts forms source-destination pairs to generate background traffic following Poisson arrivals, with the mean arriving rate varying from $1$ MBps to $12$ MBps.  We change the size of packets from $100$ Bytes to MTU (i.e., $1500$ Bytes). For each test scaneario, $50$ runs of the simulation are performed, and we obtain the simulation results by taking average of different runs.

\vspace{-0.03cm}
\subsection{Simulation results}
\vspace{-0.01cm}

\begin{figure}[htb]
\begin{center}
\epsfig{file=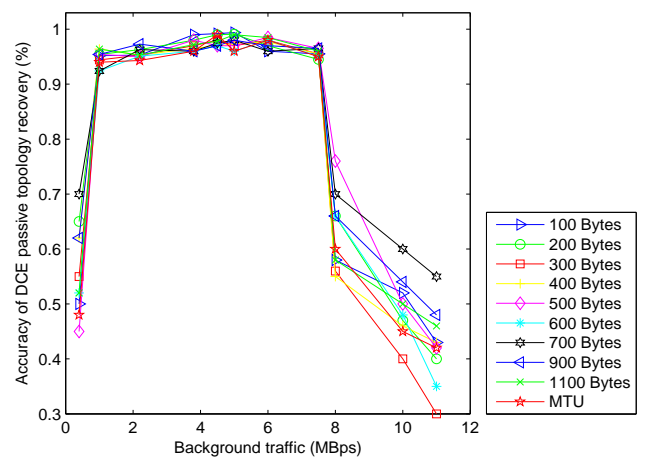, width=0.53\textwidth}
\vspace{-4mm}
\end{center}
\caption{\label{fig::accuracy_back_packsize} Topology recovery accuracy with different intensity of background traffic}
\vspace{-4mm}
\end{figure}

\begin{figure}[htb]
\begin{center}
\epsfig{file=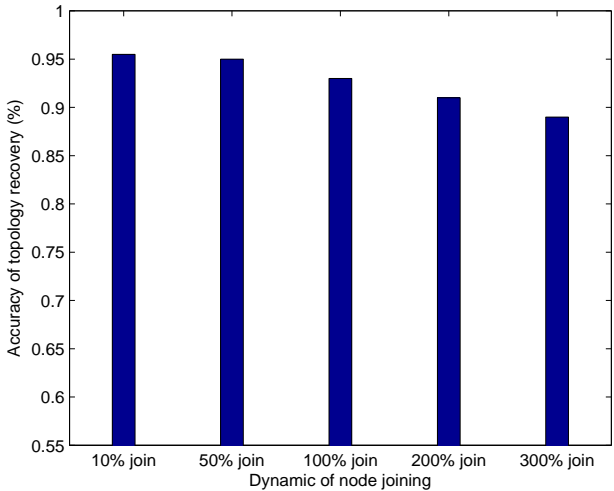, width=0.5\textwidth}
\vspace{-4mm}
\end{center}
\caption{\label{fig::result_dynamic_noNC} Performance of DCE tomography in dynamic scenarios}
\vspace{-4mm}
\end{figure}

Fig.~\ref{fig::accuracy_back_packsize} shows the topology recovery accuracy with different intensity of background traffic. The best performance ($95\%$ accuracy) is achieved when packet size is $100$ bytes and the intensity of background traffic is within the region [$1.25$ MBps,$7.5$ MBps].  When the intensity of background traffic is increased to above  $8$ MBps, the performance decreases quickly. This is because in this situation some shared paths are congested heavily, which may destroy the back-to-back property of the data packets used in DCE. It is interesting to observe that if the intensity of the background traffic is too small, e.g., below $1$ MBps, DCE tomography also shows undesirable performance. This strange phenomenon is due to the short delay on any end-to-end path, too short to be useful when we use the DCE values to infer the length of shared path. While we did not observe this phenomenon in real-world testbeds, the simulation suggests that delay correlation based methods are not effective in a lightly loaded high-speed network.

\nop{
If we change packet size from 100 Bytes to MTU the topology recovery accuracy is always desired (see other curves with different packet size in Fig.~\ref{fig::accuracy_back_packsize}).}

To test the tomography accuracy in the presence of peer join/leave, we test the following scenario: Initially the network consists of $200$ nodes, and the network is expanded as new peers join. From Fig.~\ref{fig::result_dynamic_noNC}, we can see that the performance only degrades slightly when the network becomes larger. To be specific, the recovery accuracy decreases from $95\%$ to $89\%$ when the network is expanded from $200$ nodes to $800$ nodes (i.e., $300\%$ nodes join). This indicates that the DCE passive network tomography method is robust and scalable to infer the routing topology for dynamic P2P networks. 

\section{Related work}
\label{sec::related work}

Y. Vardi was one of the first to rigorously study the problem of inferring routing topology and coined the term network tomography~\cite{Y.Var::NTESD::1996} due to the similarity between network inference and medical tomography. According to the type of data acquisition and the performance parameters of interest~\cite{RuiCastro::NTRecDeve::2004}, network tomography can be classified into (1) link-level parameter estimation based on end-to-end, path-level traffic measurements and (2) sender-receiver path-level traffic intensity estimation based on link-level traffic measurements. According to the purpose of network tomography, network tomography can also be categorized into topology discovery, link loss rates estimation, and delay tomography, etc. Based on whether or not explicit control messages are required, network tomography could be classified as active tomography~\cite{Rab::MulSouMulDesNT::2004}, \cite{CaceresR::MulInfNetInterLos::1999}, \cite{FraLo::MulInfNetInterDis::2002} and passive tomography~\cite{Y.Var::NTESD::1996}, \cite{JinCao::TimeVarNT::2000}. The former usually actively sends probing packets from the source node and analyzes measurement results at the receivers, while the latter merely utilizes the regularly transmitted data flow for further analysis.

For delay correlation measurement, Tsang et al.~\cite{YolanTsa::NetRadar::2004} develop the \emph{Network Radar} technique based on round-trip-time (RTT) measurements using \textit{TCP SYN} and \textit{SYN-ACK} segments. Nevertheless, RTT based methods may not work well, when the source-destination path and the return path are different or when the packet processing delay at the destination node is random and not negligible. 


For topology recovery, Duffield et al.~\cite{NGDuffield::NTMeaEnd2EndDelayCov::2004,NGDuffield::AdapMulTopoInf::2001} describe hierarchical clustering algorithms, which require the knowledge of all $\frac{N(N-1)}{2}$ correlation values of $N$ end nodes. The requirement implies an exhaustive probe to the network. Ni et al.~\cite{JianNi::EffDynaRouTopoInfEnd2End::2010} propose a sequential topology inference approach that needs $lN\log_lN$ (for a \emph{l-ary} tree) probing packets for the same $N$ nodes. However, when utilizing this algorithm for topology tomography, we must build a basic tree (For example, we should know \emph{l} different leaf nodes $d_1,d_2,...,d_l$ descended from a given node). This may be quite hard at the start time.

Depth-First-Search (DFS) ordering tomography introduced in~\cite{BDEriksson::ToPraNTIntTopoDiscov::2010} is closely related to our work. This approach first arranges end hosts in a DFS order then reconstructs the unknown logical topology one by one. It has been demonstrated that the total number of probing packets can be reduced from the tomography methodology in~\cite{JianNi::EffDynaRouTopoInfEnd2End::2010} by a factor of $2$. Nevertheless, the DFS ordering method does not handle problem of dynamic routing topology inference when peers join/leave the networks. 

\vspace{-0.1cm}
\section{Conclusions}
\label{sec::conclusions}
\vspace{-0.1cm}

In this paper, we propose a new delay correlation estimation (DCE) method, which is both time and bandwidth efficient and simple to implement. Based on the measurement values, we present a topology inference algorithm, which is capable of dynamically updating the routing topology when peers join/leave in P2P networks. Our DCE based method takes a free ride from the data packets transmitted in P2P networks and thus does not need any active probing messages. We implement it over real-world testbeds, including a small-scale Internet testbed and the PlanetLab. We also test its performance in large-scale networks with simulation.  The key findings of our evaluation include: 
\begin{enumerate}
	\item The DCE based passive network tomography achieves over $92\%$ accuracy in the small-scale Internet experiment and $95\%$ accuracy in the simulation for the large-scale networks.  
	\item PlanetLab may not be a proper platform to test passive network tomography, due to its virtualization techniques and usually heavily loaded end hosts.
	\item Passive network tomography does not work well in a congested network. In addition, it may not work well neither in lightly-loaded high-speed networks, particularly when end-to-end delay correlation is used as the main parameter in topology inference.   
\end{enumerate}

\section*{Acknowledgements}
This work was supported by the National Key Technology Research and Development Program of the Ministry of Science and Technology of China under Grant no. 2012BAH93F01 and the National Science Foundation of China under Grant no. 60803005.





\bibliographystyle{elsarticle-num}
\bibliography{Reference}

\begin{thebibliography}{10}
\expandafter\ifx\csname url\endcsname\relax
  \def\url#1{\texttt{#1}}\fi
\expandafter\ifx\csname urlprefix\endcsname\relax\def\urlprefix{URL }\fi
\expandafter\ifx\csname href\endcsname\relax
  \def\href#1#2{#2} \def\path#1{#1}\fi

\bibitem{Stephanos::SurP2PConDisTec::2004}
S.~Androutsellis-Theotokis, D.~Spinellis, A survey of peer-to-peer content
  distribution technologies, ACM Computing Surveys (CSUR) 36 (December 2004)
  335--371.

\bibitem{Dongyu::ModPerAnaBTP2PNet::2004}
D.~Qiu, R.~Srikant, Modeling and performance analysis of bittorrent-like
  peer-to-peer networks, ACM SIGCOMM Computer Communication Review 34 (October
  2004) 337--378.

\bibitem{Chandler::ToP2PMulShaUseGenCont::2012}
H.~Chandler, H.~Shen, L.~Zhao, J.~Stokes, J.~Li, Toward p2p-based multimedia
  sharing in user generated contents, IEEE Transactions on Parallel and
  Distributed Systems 23~(5) (May 2012) 966--975.

\bibitem{H.B::RON::2001}
D.~Andersen, H.~Balakrishnan, F.~Kaashoek, R.~Morris, Resilient overlay
  networks, in: Proceedings of SOSP, Banff, AB, Canada, Oct. 2001, pp.
  131--145.

\bibitem{R.V.B::TIAR::2003}
B.~Yao, R.~Viswanathan, F.~Chang, D.~Waddington, Topology inference in the
  presence of anonymous routers, in: Proceedings of the IEEE INFOCOM, San
  Francisco, CA, April. 2003, pp. 353--363.

\bibitem{Y.Var::NTESD::1996}
Y.~Vardi, Network tomography: estimating source-destination traffic intensities
  from link data, Journal of the American statistical association 91~(433)
  (1996) 365--377.

\bibitem{Rab::MulSouMulDesNT::2004}
M.~Rabbat, R.~Nowak, M.~Coates, Multiple source, multiple destination network
  tomography, in: Proceedings of the IEEE INFOCOM, Piscataway, NJ, USA, March
  2004, pp. 1628--1639.

\bibitem{CaceresR::MulInfNetInterLos::1999}
R.~Caceres, N.~G. Duffield, J.~Horowitz, D.~F. Towsley, Multicast-based
  inference of network internal loss characteristics, IEEE Transactions on
  Information Theory 45(7) (November 1999) 2462--2480.

\bibitem{FraLo::MulInfNetInterDis::2002}
F.~L. Presti, N.~G. Duffield, J.~Horowitz, D.~Towsley, Multicast-based
  inference of network-internal delay distributions, IEEE/ACM Transactions on
  Networking 10~(6) (December 2002) 761--775.

\bibitem{JinCao::TimeVarNT::2000}
J.~Cao, D.~Davis, S.~V. Wiel, B.~Yu, S.~Vander, W.~B. Yu, Time-varying network
  tomography: Router link data, Journal of the American statistical association
  95 (2000) 1063--1075.

\bibitem{Venkata::PassNTBayInf::2002}
V.~N. Padmanabhan, L.~Qiu, H.~J. Wang, Passive network tomography using
  bayesian inference, Microsoft Research (2002) 1--2.

\bibitem{FabioRicciato::PasTom3GNet::2006}
F.~Ricciato, F.~Vacirca, W.~Fleischer, J.~Motz, M.~Rupp, Passive tomography of
  a 3G network: Challenges and opportunities, in: Proceedings of the IEEE
  INFOCOM, 2006.

\bibitem{PasNTomErrNetNCAppr::HYao::2012}
H.~Yao, S.~Jaggi, M.~Chen, Passive network tomography for erroneous networks: A
  network coding approach, IEEE Transactions on Information Theory 58(9) (2012)
  5922--5940.

\bibitem{YolanTsa::NetRadar::2004}
Y.~Tsang, P.~Barford, R.~Nowak, Network radar: Tomography from round trip time
  measurements, in: Proceedings of the 4th ACM SIGCOMM conference on Internet
  measurement(IMC 04), 2004, pp. 175--180.

\bibitem{YolanTsa::NetDelayTomo::2003}
Y.~Tsang, R.~D. Nowak, Network delay tomography, IEEE Transactions on Signal
  Processing 51~(8) (2003) 2125--2136.

\bibitem{BDEriksson::ToPraNTIntTopoDiscov::2010}
B.~D. Eriksson, P.~Barford, R.~Nowak, Toward the practical use of network
  tomography for internet topology discovery, in: Proceedings of IEEE INFOCOM,
  2010.

\bibitem{JianNi::EffDynaRouTopoInfEnd2End::2010}
J.~Ni, H.~Xie, Tatikonda, Yang, Efficient and dynamic routing topology
  inference from end-to-end measurements, IEEE/ACM Transactions on Networking
  (2010) 123--135.

\bibitem{BitTorrent::WebSite}
Source code of bittorrent,
  \url{http://bittorrent.cvs.sourceforge.net/viewvc/bittorrent/}, the homepage
  of BitTorrent source code.

\bibitem{PlanetLab::WebSite}
Planetlab, \url{http://www.planet-lab.org/}, the homepage of PlanetLab.

\bibitem{KoungSoo::CoMon::2005}
K.~Park, V.~Pai, Comon: A monitoring infrastructure for planetlab, \url{
  http://comon.cs.princeton.edu/} (2005).

\bibitem{SteSol::ContOpeSysVir::2007}
S.~Soltesz, H.~P\"{o}tzl, M.~E. Fiuczynski, A.~Bavier, L.~Peterson,
  Container-based operating system virtualization: a scalable, high-performance
  alternative to hypervisors, in: In Proceedings of EuroSYS, 2007.

\bibitem{AndyMic::OpSySuPlanetNetSer::2004}
A.~Bavier, M.~Bowman, B.~Chun, D.~Culler, S.~Karlin, S.~Muir, L.~Peterson,
  T.~Roscoe, T.~Spalink, M.~Wawrzoniak, Operating system support for
  planetary-scale network services, in: USENIX Symposium on Networked
  Systems Design and Implementation, March 2004.

\bibitem{AcMeaSysSharEnv::JoelPaul::2007}
J.~Sommers, P.~Barford, An active measurement system for shared environments,
  in: Proceedings of the ACM IMC, October 2007, pp. 303--314.

\bibitem{OMNet++::WebSite}
Omnet++, \url{http://www.omnetpp.org/}, the homepage of OMNet++.

\bibitem{Amedina::Brite::2001}
A.~Medina, A.~Lakhina, I.~Matta, J.~Byers, Brite: An approach to universal
  topology generation, in: Proceedings of the MASCOTS, 2001.

\bibitem{RuiCastro::NTRecDeve::2004}
R.~Castro, M.~Coates, G.~Liang, R.~Nowak, B.~Yu, Network tomography:recent
  developments, Journal of Statistical Science 19~(3) (2004) 499--517.

\bibitem{NGDuffield::NTMeaEnd2EndDelayCov::2004}
N.~G. Duffield, F.~L. Presti, Network tomography from measured end-to-end delay
  covariance, IEEE/ACM Transactions On Networking 12~(6) (2004) 978--992.

\bibitem{NGDuffield::AdapMulTopoInf::2001}
N.~G. Duffield, J.~Horowitz, F.~L. Presti, Adaptive multicast topology
  inference, in: Proceedings of IEEE INFOCOM, January 2001, pp. 1636--1645.

\end{thebibliography}






\parpic{\includegraphics[width=1in,clip,keepaspectratio]{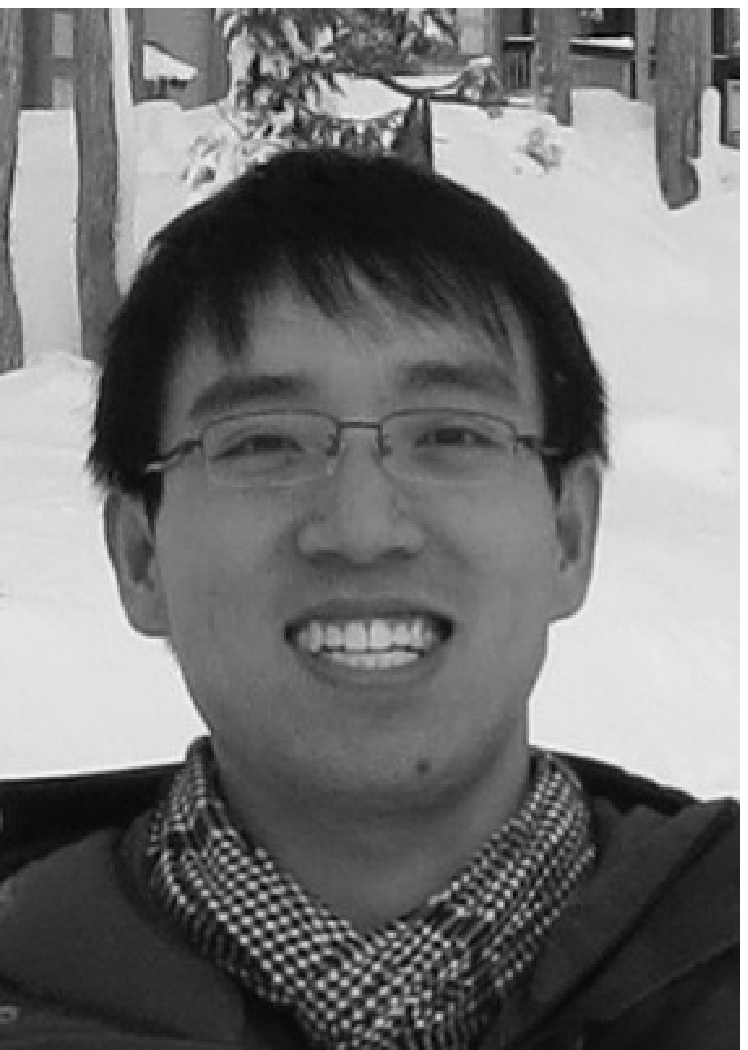}}
\noindent {\bf Peng Qin} received the B. S. degree in Electronics and Information Engineering from Huazhong University of Science and Technology, Wuhan, P. R. China, in 2009. He is currently a PhD Candidate in the Department of Electronics and Information Engineering at the Huazhong University of Science and Technology. He is now a visiting student at the University of Victoria in British Columbia, Canada. His research interests are in the areas of network tomography, network measurement, p2p network and applications of network coding.

\parpic{\includegraphics[width=1in,clip,keepaspectratio]{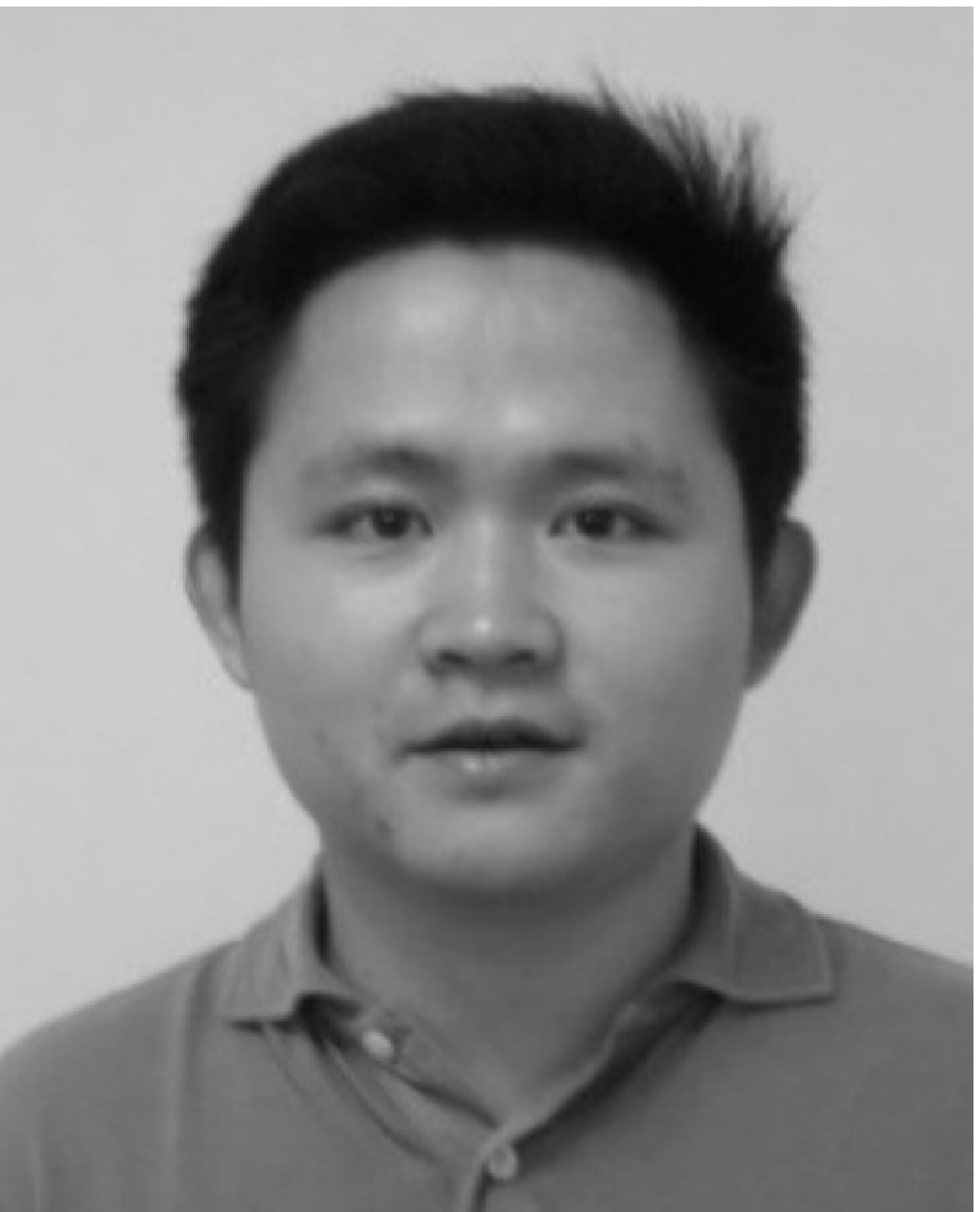}}
\noindent {\bf Bin Dai} received the B. Eng, the M. Eng degrees and the PhD degree from Huazhong University of Science and Technology of China, P. R. China in 2000, 2002 and 2006, respectively. From 2007 to 2008, he was a Research Fellow at the City University of Hong Kong. He is currently an associate professor at Department of Electronics and Information Engineering, Huazhong University of Science and Technology, P. R. China. His research interests include p2p network, wireless network, network coding, and multicast routing.

\parpic{\includegraphics[width=1in,clip,keepaspectratio]{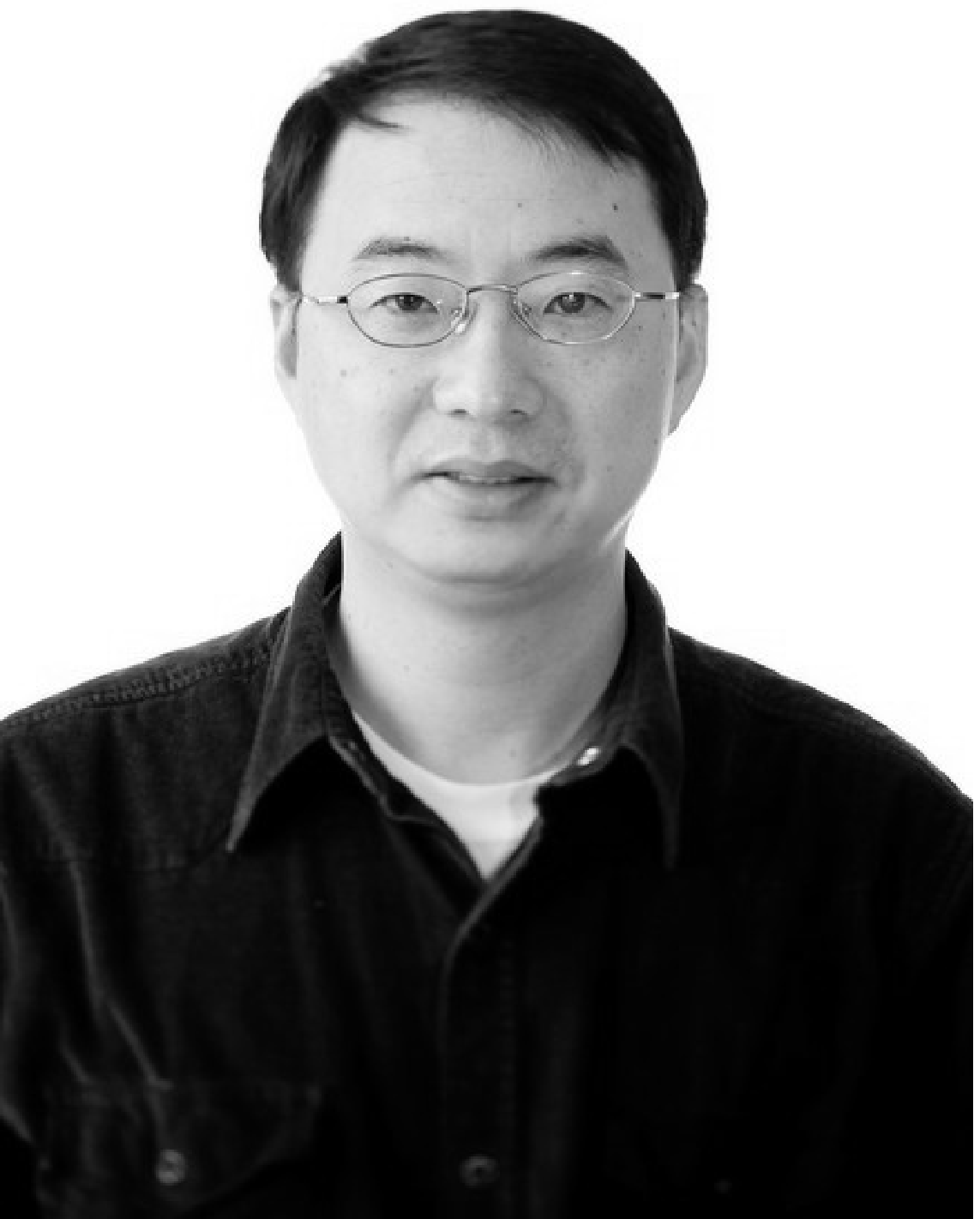}}
\noindent {\bf Kui Wu} received the PhD degree in Computer Science from the University of Alberta, Canada, in 2002. He joined the Department of Computer Science at the University of Victoria, Canada, in the same year and is currently an associate professor there. His research interests include mobile and wireless networks, network performance evaluation, and network security. He is a senior member of the IEEE.

\parpic{\includegraphics[width=1in,clip,keepaspectratio]{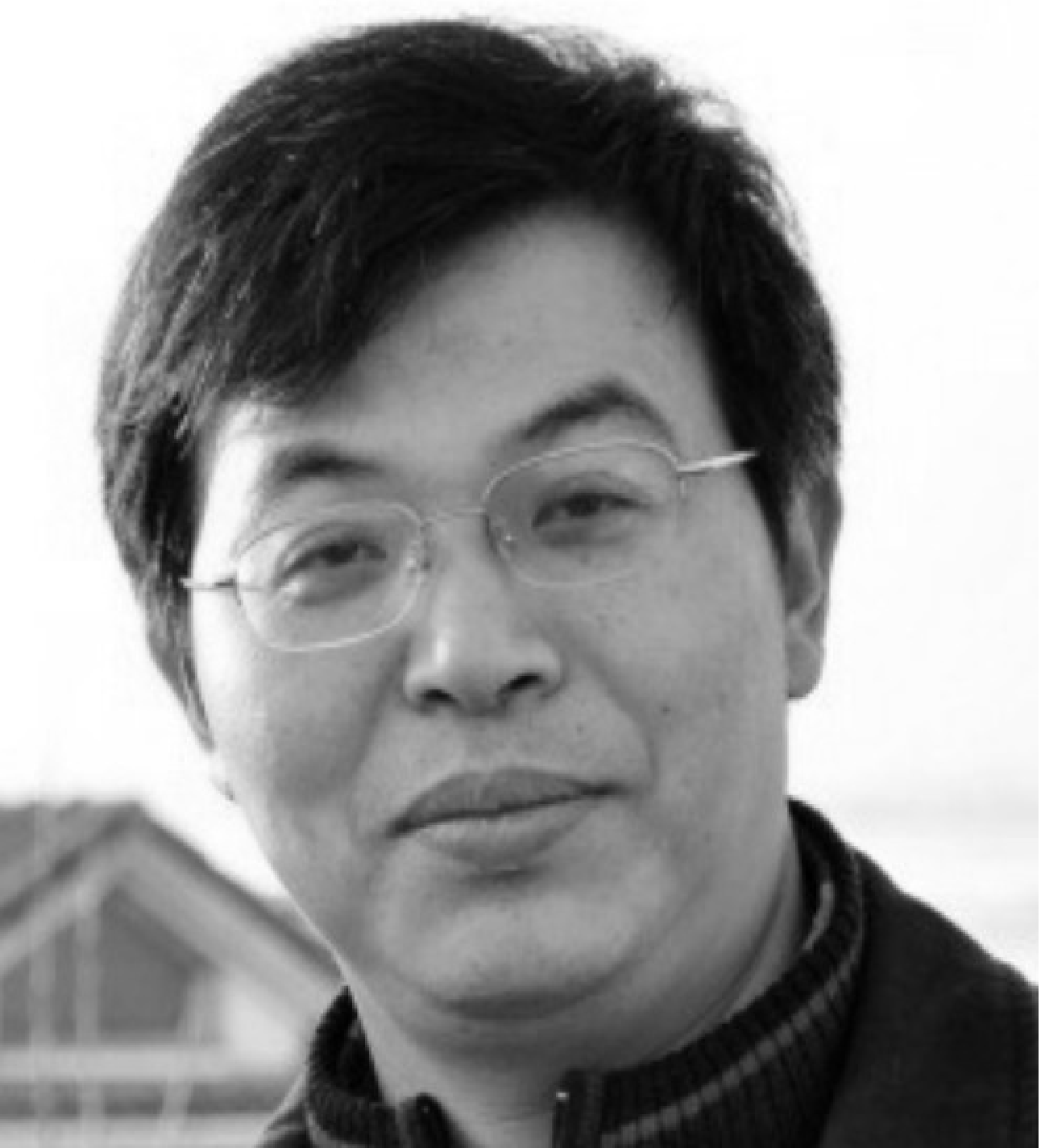}}
\noindent {\bf Benxiong Huang} received the B. S. degree in 1987 and PhD degree in 2003 from Huazhong University of Science and Technology, Wuhan, P. R. China. He is currently a professor in the Department of Electronics and Information Engineering, Huazhong University of Science and Technology, P. R. China. His research interests include next generation communication system and communication signal processing.

\parpic{\includegraphics[width=1in,clip,keepaspectratio]{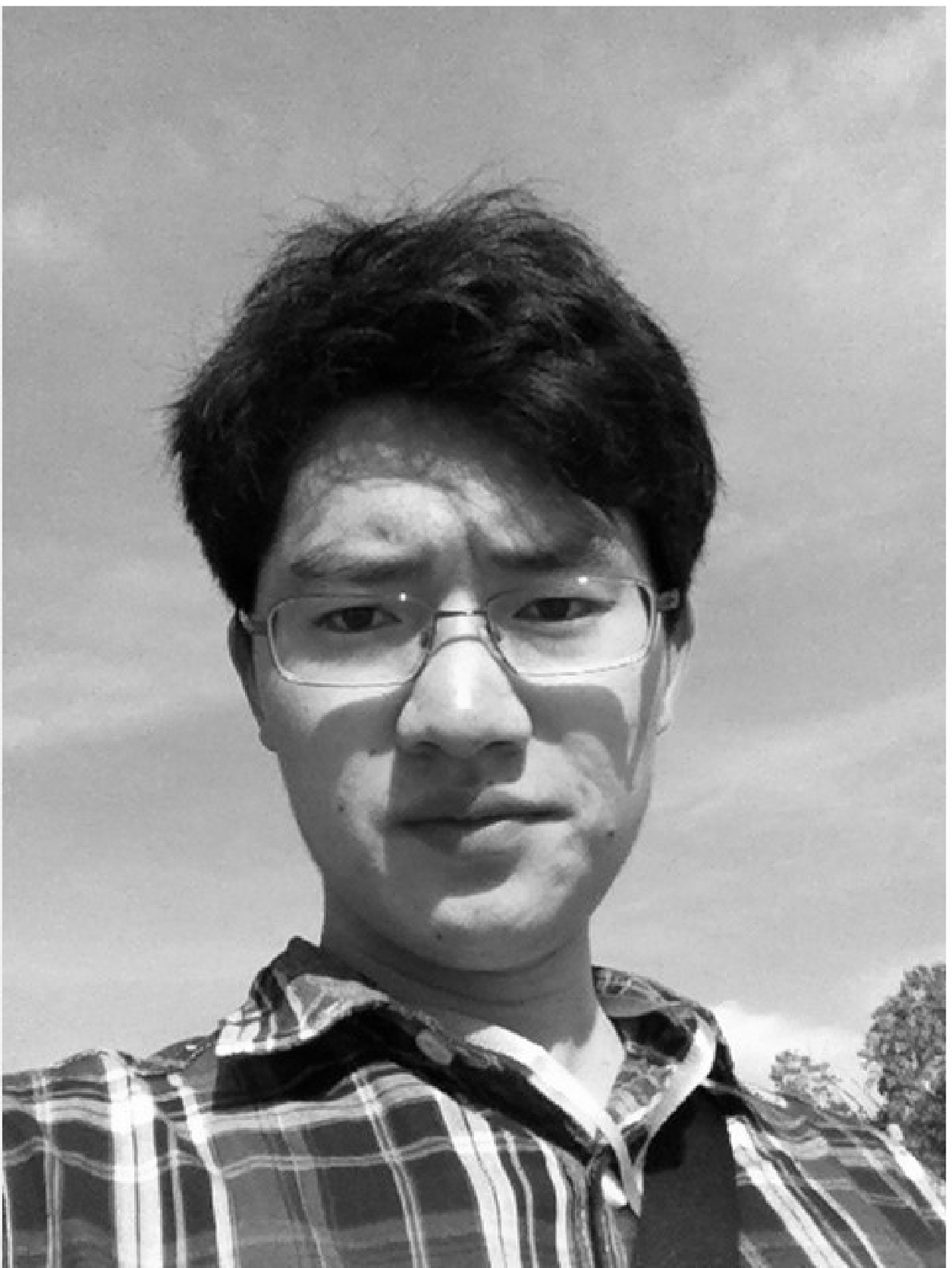}}
\noindent {\bf Guan Xu} received the B.S. degree in Electronics and Information Engineering from Huazhong University of science and technology, Wuhan, P. R. China, in 2008. He is currently a PhD Candidate in the Department of  Electronics and Information Engineering at the Huazhong University of science and technology. His research interests are in the areas of  practical network coding in P2P network, IP switch networks and SDN networks with emphasis on routing algorithms and rate control algorithms.

\end{document}